\documentclass[aps,prd,showpacs,preprintnumbers,onecolumn]{revtex4}
\usepackage{epsfig}  
\usepackage{latexsym,amsmath,amssymb,amstext}
\begin{document}  
\title{Exact chiral invariance at finite density on the lattice}  
\author{R.\ V.\ \surname{Gavai}\footnote{On sabbatical leave from Department of Theoretical Physics, Tata Institute of Fundamental Research, Homi Bhabha Road,
Mumbai 400005, India.} }  
\email{gavai@tifr.res.in}  
\affiliation{Fakult\"at f\"ur Physik, Universit\"at Bielefeld,
D-33615 Bielefeld, Germany}  
\author{Sayantan\ \surname{Sharma}}
\email{sayantan@physik.uni-bielefeld.de}  
\affiliation{Fakult\"at f\"ur Physik, Universit\"at Bielefeld,
D-33615 Bielefeld, Germany}  

\begin{abstract} 
We propose a lattice action for the overlap Dirac matrix with
nonzero chemical potential which is shown to preserve the chiral invariance on
the lattice exactly. We further demonstrate it to arise from the Domain wall by
letting the chemical potential count only the physically relevant wall modes.  
\end{abstract}
\preprint{BI-TP 2011/43, TIFR/TH/11-48}  
\maketitle  

\section{Introduction} 

Our world of strongly interacting particles has two light quarks with masses
much smaller than $\Lambda_{QCD}$, the scale of the corresponding theory,
Quantum Chromo Dynamics (QCD).  While the mass of the strange quark is
comparable to this scale, other quarks are heavy.  It has been argued on the
basis of the corresponding symmetries of the theory, called the chiral
symmetries, that QCD may have a critical point in the temperature ($T$)-baryon
number density (or equivalently, the baryonic chemical potential $\mu_B$).
Theoretical~\cite{fk,gg1,schmidt} as well as experimental searches for locating
it are currently going on~\cite{mohanty}.  Its discovery would be exciting in
many ways. Apart from becoming a new milestone in our understanding of the
nature of the strongly interacting matter, it would also be unique compared  to
the other known phase diagrams in the way theory and experiment compliment  in
locating the critical point in it.   

Due to the essentially non-perturbative nature of the problem, theoretical
efforts based on a first principles approach employ lattice QCD techniques
which have made successful predictions for many hadronic quantities such as the
heavy meson decay constants. Indeed, it has also been very successful in
application at finite temperature, leading to several interesting results for
the RHIC and other heavy ion experiments~\cite{qm11}. But its foray in the
finite density domain has been hampered by serious conceptual problems.   In
spite of the so-called `fermion sign problem', which refers to the complex
measure of the theory for nonzero chemical potential, many attempts have been
made to explore nonzero $\mu_B$, some claiming success~\cite{fk,gg1,gg2} in
locating it, while others arguing for a lack~\cite{dfp} of a critical point.
Most of these computations use the staggered quarks, primarily since the
critical point is thought to be related to chiral symmetry.  Staggered quarks
have a $U(1)_V \times U(1)_A$  chiral symmetry on the lattice, and a
corresponding order parameter, the chiral condensate.  Studying its variation
with $T$ or $\mu_B$, one can look for a chiral phase transition.  The critical
point would then just be an end point of a line of first order phase
transitions.  

Staggered quarks, however, i) break the flavour and spin symmetry on the
lattice, ii) have no flavour singlet axial symmetry and iii) have a stronger
~\cite{gss} `rooting problem' at nonzero $\mu_B$.   Since the QCD critical
point is expected~\cite{rw} to exist if only if one has two light flavours, and
the flavour singlet anomaly is mildly temperature dependent~\cite{piswil}, it
appears desirable to improve upon them.   The overlap Dirac
fermions~\cite{Neu1}, or the closely related domain wall
fermions~\cite{kaplan}, offer such a possibility to improve.  Indeed, the
overlap quarks have all the symmetries of the continuum QCD  and also have a an
index theorem~\cite{hln} as well, raising the hope that even the anomaly
effects could be well treated.   Unfortunately, adding the chemical potential
turns out to be nontrivial for them.  Bloch and Wettig~\cite{bw} made a
proposal to do so but it violates~\cite{bgs} the exact chiral invariance on
lattice as does the simple addition~\cite{gs} of a baryon number term.    In
this letter we propose an alternative which does have exact chiral invariance
on lattice for any value of the lattice spacing and any chemical potential.  It
therefore can permit, in  principle, the task of mapping out the $T$-$\mu_B$
phase diagram, assuming that  the algorithmic developments can handle the
fermion sign problem well.

\section{Formalism}  
\label{basic formalism}

The massless continuum QCD action can be written in a form where
the chiral symmetry is manifest in terms of the fields appearing in the action:
\begin{eqnarray}  
\nonumber 
S_{QCD} &=& \int d^3x~d\tau [\bar \psi {\not}D
\psi-F^{\mu \nu} F_{\mu \nu} /4] ~\\         &=& \int d^3x~d\tau [\sum_{i=L,R}
(\bar \psi_i {\not}D \psi_i)-F^{\mu \nu} F_{\mu \nu} /4] ~, 
\label{eqn:qcdlr}
\end{eqnarray} 
where $\psi_L = (1-\gamma_5)\psi/2 $ and $\psi_R = (1+\gamma_5)\psi/2 $ with $
\bar \psi_L = \bar \psi (1+\gamma_5)/2 $ and $\bar \psi_R = \bar \psi
(1-\gamma_5)/2$. The second term is the action for gluons which will play no
role in our discussion below.  We assume below that some usual convenient form
has been chosen.  Adding the canonical $\mu N$ to the first line for
investigating finite density effects is the same as adding $\mu \bar \psi_i
\gamma^4 \psi_i$ to the second line, leaving the manifest chiral symmetry
intact.  We propose that addition of chemical potential on the lattice be done
in this explicit chiral symmetry preserving manner as well, opting therefore
for the overlap quarks.

The overlap quarks have all the symmetries of the continuum QCD  but also have a
nonlocal action:   
\begin{equation} 
\label{eqn:ovelp} 
S_F = \sum_{n,m}  \bar \psi_n~a D_{ov,nm}~\psi_m  ~, 
\end{equation} 
where the sum over $n$ and $m$ runs over all the space-time lattice sites, $a$
is the lattice spacing, and the overlap Dirac matrix $D_{ov}$ is defined by
$aD_{ov}=1+\gamma_5 sgn (\gamma_5 D_W)$.  {\em sgn} denotes the sign function. 
$D_W$ is the standard Wilson-Dirac matrix on the lattice but with a negative
mass term $M\in(0,2)$:
\begin{equation} 
\label{eqn:Dwil} 
D_W(x,y) = (4-M)\delta_{x,y} - \sum_{i=1}^{4}
[U^{\dagger}_{i}(x-\hat{i})\delta_{x-\hat{i},y}\frac{1+\gamma_{i}}{2}
+\frac{1-\gamma_{i}}{2}U_{i}(x)\delta_{x+\hat{i},y}]~.
\end{equation} 
The overlap Dirac matrix satisfies Ginsparg-Wilson relation~\cite{wil},  
$\{\gamma_5, D\} = a D \gamma_5 D $ and has exact chiral symmetry on lattice. 
The corresponding infinitesimal chiral transformations~\cite{Lues} are
\begin{equation} 
\label{eqn:chrl} 
\delta \psi = i\alpha  \gamma_5(1 - \frac{a}{2}D_{ov}) \psi  ~~~{\rm and} ~~~
\delta \bar \psi =i \alpha \bar \psi (1 - \frac{a}{2}D_{ov})\gamma_5 ~.~ 
\end{equation} 
An alternate set of transformations, differing by terms of ${\cal O}(a)$ is 
\begin{equation}
\label{eqn:chrLR} 
\delta \psi = i \alpha  \gamma_5(1 - a D_{ov}) \psi  ~~~{\rm
and} ~~~   \delta \bar \psi =i \alpha \bar \psi \gamma_5 ~.~ 
\end{equation}
Since one needs to write chiral projectors on the lattice to mimic
eq.(\ref{eqn:qcdlr}), we focus on eq.(\ref{eqn:chrLR}).  The generators of the
transformation in eq.(\ref{eqn:chrLR}) satisfy $\gamma_5^2 =1$ and $ \hat
\gamma_5^2 \equiv  [\gamma_5 ( 1 - a D_{ov})]^2 =1 $.  On the other hand, the
corresponding ones in eq.(\ref{eqn:chrl}) do not square to unity.  One
therefore defines the left-right projections for quark fields as $\psi_L =
(1-\hat \gamma_5) \psi/2$ and $\psi_R = (1+ \hat \gamma_5) \psi/2$, leaving the
antiquark field decomposition as in the continuum above.  Such a decomposition
is commonly done for writing chiral gauge theories~\cite{chgu} on the lattice
and is possible since $\psi$ and $\bar \psi$ are independent fields in the
Euclidean field theory.   In analogy with the eq.(\ref{eqn:qcdlr}) of continuum
QCD, the action for the overlap quarks in presence of nonzero chemical
potential may now be written down as
\begin{eqnarray} 
S &=& \sum_n  \sum_{i=L,R} [\bar \psi_{n,i} (aD_{ov} + a\mu
\gamma^4) \psi_{n,i}]  \\  &=& \sum_n  \bar \psi_n [( 1- a\mu \gamma^4/2
)aD_{ov} + a\mu \gamma^4] \psi_n  ~. 
\label{eqn:ovelr} 
\end{eqnarray}  
It is easy to verify that i) this action is invariant under the chiral
transformation eq.(\ref{eqn:chrLR}) for all values of $a\mu$ and $a$ and ii) it
reproduces the continuum action in the limit of $a \to 0$ with $\mu \to \mu/M$
scaling.  In order to obtain the order parameter for checking if the symmetry
is spontaneously broken, one usually adds a linear breaking term.  Adding a
quark mass term, $am (\bar \psi_R \psi_L + \bar \psi_L \psi_R)$, one obtains
the order parameter valid for all $T$ and $\mu$ on the lattice by taking a
derivative of the log of the partition function with respect to $am$ as, 
\begin{equation} 
\langle \bar \psi \psi \rangle = 
\lim_{m \to 0} \lim_{V \to \infty}  \langle {\rm Tr~} {\frac{(1-a D_{ov}/2)} 
{[ a D_{ov} + (am + a\mu \gamma^4) (1 -a D_{ov}/2)]} } \rangle ~. 
\end{equation}  
The only real eigenvalues of $ a D_{ov}$ are 0 or 2,
with only the former  contributing to the order parameter.  Defining a matrix
$K_{ov} =  D_{ov}(1 -a D_{ov}/2)^{-1} $, such that $\{\gamma_5, K_{ov} \} = 0$,
the order parameter can be written in a form more analogous to the continuum :
\begin{equation} 
\langle \bar \psi \psi \rangle = \langle {\rm Tr~} {\frac {1} { a K_{ov} + am +
a\mu \gamma^4}  } \rangle~. 
\end{equation} 
Although the discussion above is for a single flavour of quark, i.e, $U(1)_L
\times U(1)_R$ symmetry, its generalization to $N_f$ flavours is
straightforward.  Indeed, since it relies only on the spin-structure, the
flavour index as well as the corresponding generator matrices just carry
through.   

\section{Domain  Wall Fermions} 
\label{dwf}  

The action in eq.(\ref{eqn:ovelr}) was obtained by demanding the left-right
symmetry to be explicit.  A physically more intuitive way to arrive it is in
the domain wall formalism.  Since only the massless domain wall modes are
physical although there are many  massive unphysical modes, the appropriate way
to introduce $\mu$ is as a Lagrange multiplier for the number of these massless
modes.    The domain wall action of ~\cite{shamir} then for nonzero chemical
potential, $\mu$, is 

\begin{eqnarray}
\label{eqn:dwacmu} 
S &=& \sum_{x,x'}\sum_{s,s'=1}^{N_5} a^4\bar \psi(x,s)\left[- \delta_{x,x'}
\left(P_-\delta_{s',s+1}+P_+\delta_{s',s-1}\right)
+  (\frac{a_5}{a}D_W(x,x')+ \delta_{x,x'})\delta_{s,s'} \right. \\ \nonumber
&+& \left.  a \mu~\gamma_4~
\delta_{x,x'}\left(\delta_{s,1}\delta_{s',1}P_-+
P_+\delta_{s,N_5}\delta_{s',N_5}\right) + am~ 
\delta_{x,x'}\left(\delta_{s,1}\delta_{s',N_5}P_+ +
P_-\delta_{s,N_5}\delta_{s',1}\right)
\right]\psi(x',s')~, 
\end{eqnarray}
where $P_\pm = (1 \pm \gamma_5)/2$ and $N_5$ and $a_5$ are
the number of sites and the lattice spacing in the fifth direction
respectively.  The physically relevant 4D massless fermion field is
identified  with the fermion fields at the boundaries of the fifth dimension
as, 
\begin{equation} 
\psi=P_- \psi_1+P_+ \psi_{N_5}~~,~~\bar\psi=\bar \psi_1 P_+ 
+\bar \psi_{N_5}P_-~. 
\end{equation}

It is convenient to visualize the five dimensional
action in terms of the  fields $\eta_i$ localized on four dimensional
branes existing at each  site $i$ along the fifth dimension, as in~\cite{eh}.  
Defining a transfer matrix, $T =(1+a_5H_W P_+)^{-1}(1-a_5 H_WP_-)$,
between pairs of neighboring branes, where $H_W=\gamma_5 D_W$, 
the five dimensional action can be rewritten in terms of these fields  as,
\begin{eqnarray} 
\nonumber  S&=&\sum_{x,s}\left[\bar\eta_1\left(P_- -ma P_+
+ a\mu\left (a_5 H_W P_- -1\right)^{-1}\gamma_4 P_-\right)\eta_1-
\bar\eta_{N_5}T^{-1}\left(P_+ -ma P_- - a\mu\left(a_5 H_W P_+ +1\right)^{-1}
\gamma_4 P_+\right)\eta_1\right.\\ &-&\left. \bar\eta_1 T^{-1}
\eta_2+\sum_{s=2}^{N_5-1} \left(\bar\eta_s \eta_s-\bar\eta_s
T^{-1}\eta_{s+1}\right)+\bar\eta_{N_5} \eta_{N_5}\right]~. 
\end{eqnarray} 
The fermion fields $\eta_i$ can be integrated out successively, resulting in a
partition function  of the form, 
\begin{equation} \label{eqn:dw5d} 
\mathcal{Z}_{5D} \equiv \int\mathcal{D} U \rm{e}^{-S_G}~J~\det D^{(5)}(ma, 
a\mu) ~,~
\end{equation}
where $J$ is the Jacobian for the transformation from $\psi$-fields  to
$\eta$-fields and the five dimensional determinant $D^{(5)}$ is given by
\begin{equation} 
\mathcal{D}^{(5)} = \det \left[P_- -ma P_+
+ a\mu \left(a_5 H_W P_- -1\right)^{-1} \gamma_4 P_- -T^{-N_5} \left(P_+ -ma
P_- - a\mu\left(a_5 H_W P_+ +1\right)^{-1}\right)\gamma_4 P_+\right].
\end{equation}
In order to obtain the overlap  matrix, the contribution of 
the bulk five dimensional modes needs to be removed from  the partition 
function.  Following ~\cite{Neu2}, we introduce
pseudo-fermions and obtain the partition function of interest as
\begin{equation}
\label{eqn:ovpart} 
\mathcal{Z}_{QCD}(T,\mu) =\int\mathcal{D} U~\rm{e}^{-S_G}~\frac{\det 
D^{(5)}(ma, a\mu)} {\det D^{(5)}(ma=1, a\mu=0)} 
\end{equation} 
where $\det D^{(5)}(ma=1, a\mu=0)$ is the contribution from the 
pseudo-fermions.  Taking first $a_5\rightarrow 0$ limit and then 
$N_5\rightarrow \infty$, the ratios of determinants turns out to be  
$ \det [D_{ov} + (1- D_{ov}/2) (ma +  a \mu \gamma^4 ) ]$, where the
dimensional $\mu$ and $m$ have been scaled by a factor of $M$ to
match with the continuum limit.   A little algebra shows that $\gamma^4$ 
can be commuted through in the determinant above to yield the same overlap 
matrix of eq.(\ref{eqn:ovelr}) with exact chiral symmetry on the lattice.

\section{Discussion}  
\label{dscn}    
The action in eq.(\ref{eqn:ovelr}) leads to an overlap fermion determinant
which is identical to that in the recent work~\cite{ns} with fermionic sources
in the overlap formalism of ~\cite{NeuNar}.  The main difference is, however,
in the necessity of sources in ~\cite{ns} to define the chiral symmetry.
Indeed, the chiral symmetry transformation there is local, defined as rotation
of the sources while our eq.  (\ref{eqn:chrLR}) is nonlocal, defined as the
rotation of quark fields.  The left-right symmetry is in-built in the formalism
there whereas we needed to introduce the left-right projections in form of $L$-
and $R$-fields to do so.    Our new fermion matrix is linear in $\mu_B$,
similar to an earlier proposal by us~\cite{gs}.   This leads us to expect it to
have the same good as well as bad properties.  In particular, only its first
derivative with respect to $a\mu_B$ is nonzero, all the rest being zero.  As a
consequence, the coefficients of the Taylor expansion of the baryonic
susceptibility in $\mu_B$, needed in estimating the location of the QCD
critical point simplify considerably.  On the other hand, the corresponding
free theory itself has a $\mu_B^2$-divergence in the baryonic susceptibility in
the continuum limit.  It is easy to see that the successful
prescription~\cite{hk} for local actions of introducing $\mu_B$ as the fourth
component of a constant Abelian gauge field to remove the divergence, will not
work in this case since the nonlocal overlap Dirac matrix intertwines all four
momentum components in a covariant manner.  We have recently shown in case of
the staggered fermions that the free theory divergence can be subtracted
~\cite{gs1} out successfully.  In particular, the resultant ratios of the
Taylor coefficients appear to be in good agreement with those where the
subtraction is effected analytically by a change of action.   While further
investigations of the finite cut-off and finite volume effects are needed to
establish it firmly, it may be hoped that a similar subtraction scheme will
work for our above overlap quarks as well.  Of course, it would be clearly
desirable to modify even the action (\ref{eqn:ovelr}), without loss of its
chiral symmetry, such that it has no $\mu_B^2$-divergences.     

\section{Acknowledgements}
\label{ackn}  

We gratefully acknowledge the financial support by the Alexander von  Humboldt
foundation and the kind hospitality of the theoretical physics group of
Bielefeld University, especially that of Frithjof Karsch and Helmut Satz.   We
thank Rajamani Narayanan for useful discussions.

\end{document}